\documentclass[apjl]{emulateapj}
\usepackage{comment}
\usepackage{ifthen}
\usepackage{lineno}

\newcommand{\ctbd}[1]{}

\newcommand{\lc}{light curve}
\newcommand{\lcs}{light curves}
\newcommand{\Lc}{Light curve}
\newcommand{\Lcs}{Light curves}

\newcommand{\band}[1]{\ensuremath{#1}-band}

\newcommand{\chisq}{\ensuremath{\chi^2}}

\newcommand{\kms}{\ensuremath{\rm km\,s^{-1}}}
\newcommand{\ms}{\ensuremath{\rm m\,s^{-1}}}

\newcommand{\gcmc}{\ensuremath{\rm g\,cm^{-3}}}
\newcommand{\ergscmsq}{\ensuremath{\rm erg\,s^{-1}\,cm^{-2}}}

\newcommand{\teff}{\ensuremath{T_{\rm eff}}}
\newcommand{\logg}{\ensuremath{\log{g}}}
\newcommand{\vsini}{\ensuremath{v \sin{i}}}
\newcommand{\feh}{\ensuremath{\rm [Fe/H]}}

\newcommand{\rsun}{\ensuremath{R_\sun}}
\newcommand{\msun}{\ensuremath{M_\sun}}
\newcommand{\lsun}{\ensuremath{L_\sun}}

\newcommand{\rstar}{\ensuremath{R_\star}}
\newcommand{\mstar}{\ensuremath{M_\star}}
\newcommand{\lstar}{\ensuremath{L_\star}}

\newcommand{\teffstar}{\ensuremath{T_{\rm eff\star}}}

\newcommand{\loggstar}{\ensuremath{\log{g_{\star}}}}

\newcommand{\rpl}{\ensuremath{R_{p}}}
\newcommand{\mpl}{\ensuremath{M_{p}}}

\newcommand{\rhopl}{\ensuremath{\rho_{p}}}

\newcommand{\arstar}{\ensuremath{a/\rstar}}
\newcommand{\zrstar}{\ensuremath{\zeta/\rstar}}

\newcommand{\rjup}{\ensuremath{R_{\rm J}}}
\newcommand{\mjup}{\ensuremath{M_{\rm J}}}

\newcommand{\pack}[1]{\textsc{\lowercase{#1}}}

\newcommand{\daophot}{\pack{daophot}}

\newcommand{\reffigl}[1]{Figure~\ref{fig:#1}}
\newcommand{\refsecl}[1]{\mbox{Section \ref{sec:#1}}}

\newcommand{\reftabl}[1]{Table~\ref{tab:#1}}

\newcommand{\Hj}{Hot Jupiter}
\newcommand{\hj}{hot Jupiter}

\newcommand{\hatcurCCra}{\ensuremath{20^{\mathrm h}49^{\mathrm m}49.80{\mathrm s}}}                                  
\newcommand{\hatcurCCdec}{\ensuremath{-24{\arcdeg}25{\arcmin}43.7{\arcsec}}}                                 
\newcommand{\hatcurCCtwomass}{2MASS~20494978-2425436}                  
\newcommand{\hatcurCCgsc}{GSC~6926-00454}                              
\newcommand{\hatcurCCapassmVshort}{\ensuremath{12.4}}                
\newcommand{\hatcurCCapassmV}{\ensuremath{12.44\pm0.2}}                
\newcommand{\hatcurCCapassmB}{\ensuremath{12.46\pm0.18}}               
\newcommand{\hatcurCCtwomassJmag}{\ensuremath{10.980\pm0.021}}         
\newcommand{\hatcurCCtwomassHmag}{\ensuremath{10.719\pm0.024}}         
\newcommand{\hatcurCCtwomassKmag}{\ensuremath{10.694\pm0.021}}         
\newcommand{\hatcurLCrprstar}{\ensuremath{0.1011\pm0.0006}}            
\newcommand{\hatcurLCimp}{\ensuremath{0.497_{-0.027}^{+0.024}}}        
\newcommand{\hatcurLCzeta}{\ensuremath{15.16\pm0.03}}                  
\newcommand{\hatcurLCdur}{\ensuremath{0.1494\pm0.0007}}                
\newcommand{\hatcurLCingdur}{\ensuremath{0.0177\pm0.0007}}             
\newcommand{\hatcurLCP}{\ensuremath{3.547851\pm0.000005}}              
\newcommand{\hatcurLCPprec}{\ensuremath{3.5478513}}                    
\newcommand{\hatcurLCPshort}{\ensuremath{3.5479}}                      
\newcommand{\hatcurLCT}{\ensuremath{2456155.96734\pm0.00014}}          
\newcommand{\hatcurSMEiteff}{\ensuremath{6325\pm83}}                   
\newcommand{\hatcurSMEizfeh}{\ensuremath{-0.139\pm0.09}}               
\newcommand{\hatcurSMEizfehshort}{\ensuremath{-0.139}}                 
\newcommand{\hatcurSMEilogg}{\ensuremath{3.92\pm0.14}}                 
\newcommand{\hatcurSMEivsin}{\ensuremath{8.97\pm1.8}}                  
\newcommand{\hatcurSMEivmac}{\ensuremath{NULL}}                        
\newcommand{\hatcurSMEivmic}{\ensuremath{NULL}}                        
\newcommand{\hatcurSMEiiteff}{\ensuremath{6351\pm76}}                  
\newcommand{\hatcurSMEiizfeh}{\ensuremath{-0.157\pm0.07}}              
\newcommand{\hatcurSMEiizfehshort}{\ensuremath{-0.157}}                
\newcommand{\hatcurSMEiilogg}{\ensuremath{4.23\pm0.0}}                 
\newcommand{\hatcurSMEiivsin}{\ensuremath{9.12\pm1.31}}                
\newcommand{\hatcurSMEiivmac}{\ensuremath{NULL}}                       
\newcommand{\hatcurSMEiivmic}{\ensuremath{NULL}}                       
\newcommand{\hatcurLBii}{\ensuremath{0.1903}}                          
\newcommand{\hatcurLBiii}{\ensuremath{0.3599}}                         
\newcommand{\hatcurLBig}{\ensuremath{0.4135}}                          
\newcommand{\hatcurLBiig}{\ensuremath{0.3301}}                         
\newcommand{\hatcurLBir}{\ensuremath{0.2592}}                          
\newcommand{\hatcurLBiir}{\ensuremath{0.3725}}                         
\newcommand{\hatcurISOmshort}{\ensuremath{1.21}}                       
\newcommand{\hatcurISOmlong}{\ensuremath{1.209\pm0.036}}               
\newcommand{\hatcurISOrlong}{\ensuremath{1.404\pm0.030}}               
\newcommand{\hatcurISOlogg}{\ensuremath{4.22\pm0.01}}                  
\newcommand{\hatcurISOlum}{\ensuremath{2.86\pm0.21}}                   
\newcommand{\hatcurISOmv}{\ensuremath{3.62\pm0.09}}                    
\newcommand{\hatcurISOage}{\ensuremath{3.2_{-0.4}^{+0.6}}}             
\newcommand{\hatcurISOMK}{\ensuremath{2.46\pm0.05}}                    
\newcommand{\hatcurISOspec}{F}                                         
\newcommand{\hatcurRVK}{\ensuremath{125.7\pm15.7}}                     
\newcommand{\hatcurRVjitterA}{\ensuremath{51.0}}                       
\newcommand{\hatcurRVjitterB}{\ensuremath{9.5}}                        
\newcommand{\hatcurRVjitterC}{\ensuremath{65.0}}                       
\newcommand{\hatcurPPi}{\ensuremath{86.2\pm0.3}}                       
\newcommand{\hatcurPPlogg}{\ensuremath{3.14\pm0.06}}                   
\newcommand{\hatcurPPar}{\ensuremath{7.42\pm0.12}}                     
\newcommand{\hatcurPParel}{\ensuremath{0.0485_{-0.0006}^{+0.0004}}}    
\newcommand{\hatcurPPrho}{\ensuremath{0.51\pm0.07}}                    
\newcommand{\hatcurPPmshort}{\ensuremath{1.07}}                        
\newcommand{\hatcurPPmlong}{\ensuremath{1.071\pm0.136}}                
\newcommand{\hatcurPPrshort}{\ensuremath{1.38}}                        
\newcommand{\hatcurPPrlong}{\ensuremath{1.381\pm0.035}}                
\newcommand{\hatcurPPmrcorr}{\ensuremath{0.15}}                        
\newcommand{\hatcurPPteff}{\ensuremath{1648\pm24}}                     
\newcommand{\hatcurPPtheta}{\ensuremath{0.062\pm0.008}}                
\newcommand{\hatcurPPfluxavg}{\ensuremath{1.66\pm0.10}}                
\newcommand{\hatcurPPfluxavgdim}{\ensuremath{9}}                       
\newcommand{\hatcurXdist}{\ensuremath{453\pm11}}                       
\newcommand{\hatcurXEBV}{\ensuremath{0.098\pm0.054}}                   
\newcommand{\hatcurCCapassmVeccen}{\ensuremath{12.44\pm0.2}}           
\newcommand{\hatcurCCapassmBeccen}{\ensuremath{12.46\pm0.18}}          
\newcommand{\hatcurCCtwomassJmageccen}{\ensuremath{10.980\pm0.021}}    
\newcommand{\hatcurCCtwomassHmageccen}{\ensuremath{10.719\pm0.024}}    
\newcommand{\hatcurCCtwomassKmageccen}{\ensuremath{10.694\pm0.021}}    
\newcommand{\hatcurLCrprstareccen}{\ensuremath{0.1010\pm0.0005}}       
\newcommand{\hatcurLCimpeccen}{\ensuremath{0.489_{-0.029}^{+0.024}}}   
\newcommand{\hatcurLCzetaeccen}{\ensuremath{15.22_{-0.06}^{+0.11}}}    
\newcommand{\hatcurLCdureccen}{\ensuremath{0.1486\pm0.0011}}           
\newcommand{\hatcurLCingdureccen}{\ensuremath{0.0175\pm0.0007}}        
\newcommand{\hatcurLCPeccen}{\ensuremath{3.547849\pm0.000005}}         
\newcommand{\hatcurLCTeccen}{\ensuremath{2456092.10547\pm0.00039}}     
\newcommand{\hatcurSMEiteffeccen}{\ensuremath{6325\pm83}}              
\newcommand{\hatcurSMEizfeheccen}{\ensuremath{-0.139\pm0.09}}          
\newcommand{\hatcurSMEizfehshorteccen}{\ensuremath{-0.139}}            
\newcommand{\hatcurSMEiloggeccen}{\ensuremath{3.92\pm0.14}}            
\newcommand{\hatcurSMEivsineccen}{\ensuremath{8.97\pm1.8}}             
\newcommand{\hatcurSMEivmaceccen}{\ensuremath{NULL}}                   
\newcommand{\hatcurSMEivmiceccen}{\ensuremath{NULL}}                   
\newcommand{\hatcurSMEiiteffeccen}{\ensuremath{6338\pm76}}             
\newcommand{\hatcurSMEiizfeheccen}{\ensuremath{-0.148\pm0.07}}         
\newcommand{\hatcurSMEiizfehshorteccen}{\ensuremath{-0.148}}           
\newcommand{\hatcurSMEiiloggeccen}{\ensuremath{4.08\pm0.0}}            
\newcommand{\hatcurSMEiivsineccen}{\ensuremath{9.05\pm1.31}}           
\newcommand{\hatcurSMEiivmaceccen}{\ensuremath{NULL}}                  
\newcommand{\hatcurSMEiivmiceccen}{\ensuremath{NULL}}                  
\newcommand{\hatcurLBiieccen}{\ensuremath{0.1915}}                     
\newcommand{\hatcurLBiiieccen}{\ensuremath{0.3597}}                    
\newcommand{\hatcurLBigeccen}{\ensuremath{0.4160}}                     
\newcommand{\hatcurLBiigeccen}{\ensuremath{0.3286}}                    
\newcommand{\hatcurLBireccen}{\ensuremath{0.2607}}                     
\newcommand{\hatcurLBiireccen}{\ensuremath{0.3721}}                    
\newcommand{\hatcurISOmlongeccen}{\ensuremath{1.300\pm0.088}}          
\newcommand{\hatcurISOrlongeccen}{\ensuremath{1.725\pm0.242}}          
\newcommand{\hatcurISOloggeccen}{\ensuremath{4.08\pm0.10}}             
\newcommand{\hatcurISOlumeccen}{\ensuremath{4.28_{-1.02}^{+1.53}}}     
\newcommand{\hatcurISOmveccen}{\ensuremath{3.19\pm0.31}}               
\newcommand{\hatcurISOageeccen}{\ensuremath{3.1_{-0.4}^{+0.8}}}        
\newcommand{\hatcurISOMKeccen}{\ensuremath{2.01\pm0.30}}               
\newcommand{\hatcurRVKeccen}{\ensuremath{131.7\pm17.2}}                
\newcommand{\hatcurRVrkeccen}{\ensuremath{-0.355_{-0.089}^{+0.122}}}   
\newcommand{\hatcurRVrheccen}{\ensuremath{0.356_{-0.252}^{+0.101}}}    
\newcommand{\hatcurRVkeccen}{\ensuremath{-0.176\pm0.080}}              
\newcommand{\hatcurRVheccen}{\ensuremath{0.175\pm0.115}}               
\newcommand{\hatcurRVjitterAeccen}{\ensuremath{51.0}}                  
\newcommand{\hatcurRVjitterBeccen}{\ensuremath{9.5}}                   
\newcommand{\hatcurRVjitterCeccen}{\ensuremath{65.0}}                  
\newcommand{\hatcurRVecceneccen}{\ensuremath{0.253\pm0.110}}           
\newcommand{\hatcurRVomegaeccen}{\ensuremath{134\pm25}}                
\newcommand{\hatcurPPieccen}{\ensuremath{84.3_{-2.0}^{+1.3}}}          
\newcommand{\hatcurPPloggeccen}{\ensuremath{2.99\pm0.12}}              
\newcommand{\hatcurPPareccen}{\ensuremath{6.18_{-0.64}^{+0.88}}}       
\newcommand{\hatcurPPareleccen}{\ensuremath{0.0497\pm0.0011}}          
\newcommand{\hatcurPPrhoeccen}{\ensuremath{0.28_{-0.08}^{+0.18}}}      
\newcommand{\hatcurPPmlongeccen}{\ensuremath{1.138\pm0.152}}           
\newcommand{\hatcurPPrlongeccen}{\ensuremath{1.697\pm0.238}}           
\newcommand{\hatcurPPmrcorreccen}{\ensuremath{0.24}}                   
\newcommand{\hatcurPPteffeccen}{\ensuremath{1816\pm127}}               
\newcommand{\hatcurPPthetaeccen}{\ensuremath{0.051_{-0.008}^{+0.012}}} 
\newcommand{\hatcurPPfluxavgeccen}{\ensuremath{2.46_{-0.55}^{+0.90}}}  
\newcommand{\hatcurXdisteccen}{\ensuremath{556\pm78}}                  
\newcommand{\hatcurXEBVeccen}{\ensuremath{0.096\pm0.053}}              

\newcommand{\hatcur}{HATS-3}
\newcommand{\hatcurb}{HATS-3b}

\newcommand{\hatcurSMEversion}{ii}                                       
\newcommand{\hatcurSMEteff}{\ifthenelse{\equal{\hatcurSMEversion}{i}}{\hatcurSMEiteff}{\hatcurSMEiiteff}}
\newcommand{\hatcurSMEzfeh}{\ifthenelse{\equal{\hatcurSMEversion}{i}}{\hatcurSMEizfeh}{\hatcurSMEiizfeh}}
\newcommand{\hatcurSMEzfehshort}{\ifthenelse{\equal{\hatcurSMEversion}{i}}{\hatcurSMEizfehshort}{\hatcurSMEiizfehshort}}
\newcommand{\hatcurSMElogg}{\ifthenelse{\equal{\hatcurSMEversion}{i}}{\hatcurSMEilogg}{\hatcurSMEiilogg}}
\newcommand{\hatcurSMEvsin}{\ifthenelse{\equal{\hatcurSMEversion}{i}}{\hatcurSMEivsin}{\hatcurSMEiivsin}}
\newcommand{\hatcurSMEvmac}{\ifthenelse{\equal{\hatcurSMEversion}{i}}{\hatcurSMEivmac}{\hatcurSMEiivmac}}
\newcommand{\hatcurSMEvmic}{\ifthenelse{\equal{\hatcurSMEversion}{i}}{\hatcurSMEivmic}{\hatcurSMEiivmic}}

\newcommand{\hatcurSMEversioneccen}{i}                                       
\newcommand{\hatcurSMEteffeccen}{\ifthenelse{\equal{\hatcurSMEversioneccen}{i}}{\hatcurSMEiteffeccen}{\hatcurSMEiiteffeccen}}
\newcommand{\hatcurSMEzfeheccen}{\ifthenelse{\equal{\hatcurSMEversioneccen}{i}}{\hatcurSMEizfeheccen}{\hatcurSMEiizfeheccen}}
\newcommand{\hatcurSMEzfehshorteccen}{\ifthenelse{\equal{\hatcurSMEversioneccen}{i}}{\hatcurSMEizfehshorteccen}{\hatcurSMEiizfehshorteccen}}
\newcommand{\hatcurSMEloggeccen}{\ifthenelse{\equal{\hatcurSMEversioneccen}{i}}{\hatcurSMEiloggeccen}{\hatcurSMEiiloggeccen}}
\newcommand{\hatcurSMEvsineccen}{\ifthenelse{\equal{\hatcurSMEversioneccen}{i}}{\hatcurSMEivsineccen}{\hatcurSMEiivsineccen}}
\newcommand{\hatcurSMEvmaceccen}{\ifthenelse{\equal{\hatcurSMEversioneccen}{i}}{\hatcurSMEivmaceccen}{\hatcurSMEiivmaceccen}}
\newcommand{\hatcurSMEvmiceccen}{\ifthenelse{\equal{\hatcurSMEversioneccen}{i}}{\hatcurSMEivmiceccen}{\hatcurSMEiivmiceccen}}

\newcommand{\hatcurisoshort}{YY}

\newcommand{\hatcurisocite}{yi:2001}
\newcommand{\hatcurlumind}{\arstar}
\newcommand{\hatcurjhkfilset}{ESO}
\newcommand{\hatrecon}{reconnaissance}
\newcommand{\hatreconcap}{Reconnaissance}

\newboolean{emulateapj}
\setboolean{emulateapj}{true}

\newboolean{rvtablelong}
\setboolean{rvtablelong}{true}

\newboolean{astroph}
\setboolean{astroph}{true}

\shortauthors{Bayliss et al.}
\shorttitle{\hatcur\lowercase{b}}
\ifthenelse{\boolean{emulateapj}}{
    
}{
    
}
\ifthenelse{\boolean{emulateapj}}{
    \newcommand{\titlestar}{$\star$}
}{
    \newcommand{\titlestar}{\star}
}
\ifthenelse{\boolean{emulateapj}}{
    \newcommand{\titlestarstar}{$\star\star$}
}{
    \newcommand{\titlestarstar}{\star\star}
}

\begin{document}


\title{\hatcur\lowercase{b}: An inflated hot Jupiter transiting an F-type star}

\author{
D.~Bayliss\altaffilmark{1},
G.~Zhou\altaffilmark{1},
K.~Penev\altaffilmark{2,3},
G.~\'A.~Bakos\altaffilmark{2,3,\titlestar,\titlestarstar},
J.~D.~Hartman\altaffilmark{2,3},
A.~Jord\'an\altaffilmark{4},
L.~Mancini\altaffilmark{5},
M.~Mohler\altaffilmark{5},
V.~Suc\altaffilmark{4},
M.~Rabus\altaffilmark{4},
B.~B\'eky\altaffilmark{3},
Z.~Csubry\altaffilmark{2,3},
L.~Buchhave\altaffilmark{6},
T.~Henning\altaffilmark{5},
N.~Nikolov\altaffilmark{5},
B.~Cs\'ak\altaffilmark{5},
R.~Brahm\altaffilmark{4},
N.~Espinoza\altaffilmark{4},
R.~W.~Noyes\altaffilmark{3},
B.~Schmidt\altaffilmark{1},
P.~Conroy\altaffilmark{1},
D.~J.~Wright,\altaffilmark{7,8},
C.~G.~Tinney,\altaffilmark{7,8},
B.~C.~Addison,\altaffilmark{7,8},
P.~D.~Sackett,\altaffilmark{1},
D.~D.~Sasselov\altaffilmark{3},
J.~L\'az\'ar\altaffilmark{9},
I.~Papp\altaffilmark{9},
P.~S\'ari\altaffilmark{9}
}
\altaffiltext{1}{Research School of Astronomy and Astrophysics, Australian National University, Canberra, ACT 2611, Australia; email: daniel@mso.anu.edu.au}

\altaffiltext{2}{Department of Astrophysical Sciences,
	Princeton University, NJ 08544, USA}

\altaffiltext{3}{Harvard-Smithsonian Center for Astrophysics,
	Cambridge, MA, USA}

\altaffiltext{$\star$}{Alfred P.~Sloan Research Fellow}

\altaffiltext{$\star\star$}{Packard Fellow}

\altaffiltext{4}{Departamento de Astronom\'ia y Astrof\'isica, Pontificia
	Universidad Cat\'olica de Chile, Av.\ Vicu\~na Mackenna 4860, 7820436 Macul,
	Santiago, Chile}

\altaffiltext{5}{Max Planck Institute for Astronomy, Heidelberg,
	Germany}

\altaffiltext{6}{Niels Bohr Institute, Copenhagen University, Denmark}

\altaffiltext{7}{Exoplanetary Science at UNSW, School of Physics,
  University of New South Wales, 2052, Australia}

\altaffiltext{8}{Australian Centre for Astrobiology, University of New South Wales,
2052, Australia}

\altaffiltext{9}{Hungarian Astronomical Association, Budapest,
	Hungary}

\altaffiltext{$\dagger$}{
The HATSouth network is operated by a collaboration consisting of
Princeton University (PU), the Max Planck Institute f\"ur Astronomie
(MPIA), and the Australian National University (ANU).  The station at
Las Campanas Observatory (LCO) of the Carnegie Institute, is operated
by PU in conjunction with collaborators at the Pontificia Universidad
Cat\'olica de Chile (PUC), the station at the High Energy Spectroscopic
Survey (HESS) site is operated in conjunction with MPIA, and the
station at Siding Spring Observatory (SSO) is operated jointly with
ANU.
}


\begin{abstract}

  We report the discovery by the HATSouth survey of \hatcurb{}, a
  transiting extrasolar planet orbiting a V=\hatcurCCapassmVshort\
  \hatcurISOspec\ dwarf star.  \hatcurb{} has a period of
  $P=\hatcurLCPshort$\,d, mass of $\mpl = \hatcurPPmshort$\,\mjup, and
  radius of $\rpl = \hatcurPPrshort$\,\rjup.  Given the radius of the
  planet, the brightness of the host star, and the stellar rotational
  velocity ($\vsini=9.0$\,\kms), this system will make an interesting
  target for future observations to measure the Rossiter-McLaughlin
  effect and determine its spin-orbit alignment.  We detail the
  low/medium-resolution \hatrecon{} spectroscopy that we are now using
  to deal with large numbers of transiting planet candidates produced by
  the HATSouth survey. We show that this important step in discovering
  planets produces \logg{} and \teff{} parameters at a precision
  suitable for efficient candidate vetting, as well as efficiently
  identifying stellar mass eclipsing binaries with radial velocity
  semi-amplitudes as low as 1\,\kms.

\setcounter{footnote}{0}
\end{abstract}



\section{Introduction}
\label{sec:introduction}
Transiting exoplanets provide us with the primary source of information
about planets outside our own Solar System.  These are the only planets
for which we can routinely and accurately measure both mass and radius.
In addition, they provide the possibility for further follow-up
observations to measure other physical properties such as brightness
temperature \citep[e.g.][]{2007Natur.447..183K}, spin-orbit alignment
\citep[e.g.][]{2000A&A...359L..13Q}, and atmospheric composition
\citep[e.g.][]{2002ApJ...568..377C}.

There are 187 transiting exoplanets with published masses and
radii\footnote{http://exoplanets.org, as of 2013 June 1}, primarily
discovered by the dedicated transit surveys of WASP
\citep{2006PASP..118.1407P}, HATNet \citep{bakos:2004:hatnet}, COROT
\citep{2009A&A...506..411A}, and Kepler \citep{2010Sci...327..977B}.  A
class of planets known as ``{\hj}s'', with short periods (P$<$10\,d) and
masses/radii similar to Jupiter, account for the majority of these
discoveries.  {\Hj}s appear to be rare, occurring at a rate of around
0.4\% around solar-type stars as determined by transit surveys
\citep{2011ApJ...743..103B, 2013ApJ...766...81F}.  This rarity, coupled
with the difficulty in detecting the $\sim$1\% transit feature from a
typical {\hj}, has meant that the task of building up a statistically
significant set of {\hj}s has progressed relatively slowly.  However the
task is important for two primary reasons.  Firstly, individual systems
can be studied in great detail to probe the nature of the exoplanet.
Secondly, global trends for giant planets which require a statistically
significant sample can be uncovered to better understand the formation
and migration of planets.

The discovery of \hatcurb{} fits into both these categories.  With a
host star magnitude of V=\hatcurCCapassmVshort\ it is a promising system for future
spectroscopic and photometric follow-up studies, while it adds to the small
set of known planets for which orbital and physical properties
have been precisely measured.     

The layout of the paper is as follows. In \refsecl{obs}, we report the
detection of the photometric signal and the follow-up spectroscopic and
photometric observations of \hatcur{}.  In \refsecl{analysis}, we
describe the analysis of the data, beginning with the determination of
the stellar parameters, continuing with a discussion of the methods
used to rule out non-planetary, false positive scenarios which could
mimic the photometric and spectroscopic observations, and finishing
with a description of our global modelling of the photometry and radial
velocities.  The discovery of \hatcurb{} is discussed in
\refsecl{discussion}, along with how it fits into the landscape of known {\hj}s.

\section{Observations}
\label{sec:obs}
\subsection{Photometric detection}
\label{sec:detection}
\ifthenelse{\boolean{emulateapj}}{
    \begin{deluxetable*}{llrrr}
}{
    \begin{deluxetable}{llrrr}
}
\tablewidth{0pc}
\tabletypesize{\scriptsize}
\tablecaption{
    Summary of photometric observations
    \label{tab:photobs}
}
\tablehead{
    \multicolumn{1}{c}{Facility}          &
    \multicolumn{1}{c}{Date(s)}             &
    \multicolumn{1}{c}{Number of Images}\tablenotemark{a}      &
    \multicolumn{1}{c}{Cadence (s)}\tablenotemark{b}         &
    \multicolumn{1}{c}{Filter}            \\
    &
    &
    &
    &
}
\startdata
\bf{Discovery}\\
HS-2 (Chile)     & 2009 Sep--2010 Sep & 5660 & 284 & Sloan~$r$ \\
HS-4 (Namibia)   & 2009 Sep--2010 Sep & 8860 & 288 & Sloan~$r$ \\
HS-6 (Australia) & 2010 Aug--2010 Sep &  198 & 288 & Sloan~$r$ \\
\\
\bf{Follow-Up}\\
FTS/Spectral & 2012 Jun 20    &  143 &  50 & Sloan~$i$ \\
FTS/Spectral & 2012 Jul 15    &  377 &  50 & Sloan~$i$ \\
MPG/ESO2.2/GROND & 2012 Aug 21    &  186 & 129 & Sloan~$g$ \\
MPG/ESO2.2/GROND & 2012 Aug 21    &  186 & 129 & Sloan~$r$ \\
MPG/ESO2.2/GROND & 2012 Aug 21    &  185 & 129 & Sloan~$i$ \\
MPG/ESO2.2/GROND & 2012 Aug 21    &  185 & 129 & Sloan~$z$ \\
[-1.5ex]
\enddata 
\tablenotetext{a}{
  Excludes images which were rejected as significant outliers in the
  fitting procedure.
}
\tablenotetext{b}{
  The mode time difference between consecutive points in each \lc.  Due to visibility, weather, pauses for focusing, etc., none
  of the \lcs{} have perfectly uniform time sampling.
}
\ifthenelse{\boolean{emulateapj}}{
    \end{deluxetable*}
}{
    \end{deluxetable}
}

\begin{figure}[!ht]
\plotone{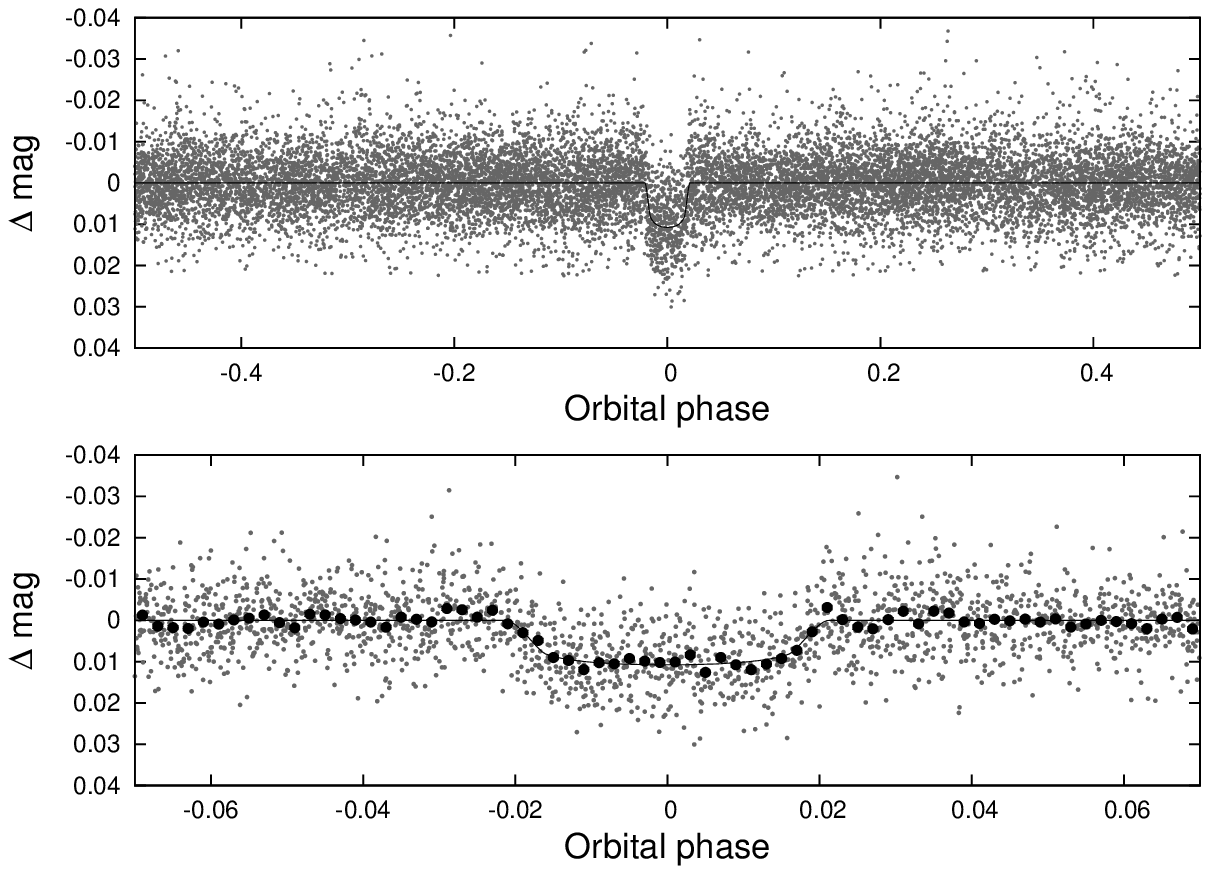}
\caption{
        The discovery \lc{} for \hatcurb{}.  The top panel is
        the unbinned, relative, instrumental \band{r} \lc{}, phase-folded to
        the period $P = \hatcurLCPprec$\,days resulting from the global 
        fit described in \refsecl{analysis}.  The lower
        panel shows a zoom-in around the transit; dark filled
        circles are the \lc{} binned at 0.002 in phase.  The solid line
        in both panels is the best-fit transit model (see
        \refsecl{analysis}).  As a result of blending and our
        noise-filtering procedure, the transit detected in the HATSouth
        \lc{} for \hatcur{} is 15\% shallower than the true
        \band{r} transit.
        \label{fig:hatsouth}}
\end{figure}

The exoplanet \hatcurb{} was first identified as a transiting exoplanet
candidate based on 14719 photometric observations of its host star
\hatcur{} (also known as \hatcurCCtwomass; $\alpha = \hatcurCCra$,
$\delta = \hatcurCCdec$; J2000), from the HATSouth global network of
automated telescopes \citep{bakos:2013:hatsouth}.  The first three
entries of \reftabl{photobs} summarize these HATSouth discovery
observations.  For this particular candidate, observations were
primarily performed by the HS2 and HS4 units (in Chile and Namibia
respectively) over the period of a year from 2009 Sept to 2010 Sept.
The HS6 unit (in Australia) only contributed a small number of images as
it was under construction and commissioning during the period when the
field containing \hatcur{} was most intensively monitored by the
HATSouth network.

Details relating to the observation, reduction, and analysis of the
HATSouth photometric discovery data are fully described in
\citet{bakos:2013:hatsouth}.  Here we provide a brief summary of the
salient points.

The HATSouth observations consist of four-minute \band{r} exposures
produced using 24 Takahashi E180 astrographs (18cm diameter primary
mirrors) coupled to Apogee 4K$\times$4K U16M Alta CCDs.  Photometry is
performed using an aperture photometry pipeline and \lcs{} are
detrended using External Parameter Decorrelation
\citep[EPD;][]{bakos:2010:hat11} and the Trend Filtering Algorithm (TFA)
of \citet{kovacs:2005:TFA}.  \Lcs{} are searched for
transit events using an implementation of the Box-fitting Least Squares
algorithm \citep[BLS;][]{kovacs:2002:BLS}.

We detected a significant transit signal
in the \lc{} of \hatcur{} (see \reffigl{hatsouth}).  Based on this
detection we initiated the multiphase follow-up procedure detailed below.

\subsection{\hatreconcap{} Spectroscopy}
\label{sec:reconspec}
The HATSouth global network of telescopes produces well over 100
candidates each year.  To efficiently follow-up these candidates, we
undertake a series of \hatrecon{} spectroscopic observations before
attempting high-resolution spectroscopy.  These \hatrecon{} observations
consist of spectral typing candidates (Section~\ref{sec:wifesspec}) and
medium-resolution radial velocities (Section~\ref{sec:wifesrv}).  We
summarize the \hatrecon{} spectroscopy observations taken for \hatcur{}
in \reftabl{specobssummary}.

\subsubsection{\hatreconcap{} Spectral Typing}
\label{sec:wifesspec}
The aim of spectral classification during \hatrecon{} spectroscopy is to
1) efficiently identify and reject candidate host stars that are giants
and therefore inconsistent (in our photometric regime) with the
planet-star scenario, and 2) determine stellar parameters so that we can
identify interesting transiting systems and prioritise the follow-up
observations. For this purpose, a low-resolution spectrum is typically
the first step in following up a HATSouth candidate.

A spectrum of \hatcur{} was
obtained using the Wide Field Spectrograph \citep[WiFeS;][]{2007Ap&amp;SS.310..255D} on the ANU 2.3\,m telescope on 2012 April
10. WiFeS is a dual-arm image slicer integral field spectrograph.  For spectral typing we use the blue arm with the B3000 grating, which
delivers a resolution of R=$\lambda/\Delta \lambda$=3000 from 3500--6000\,\AA.  Flux calibrations are performed according to
\citet{1999PASP..111.1426B} using spectrophotometric standard stars from
\citet{1992PASP..104..533H} and \citet{1999PASP..111.1426B}. A full description of
the instrument configurations can be found in \citet{penev:2013:hats1}.

\begin{figure}[!ht]
\plotone{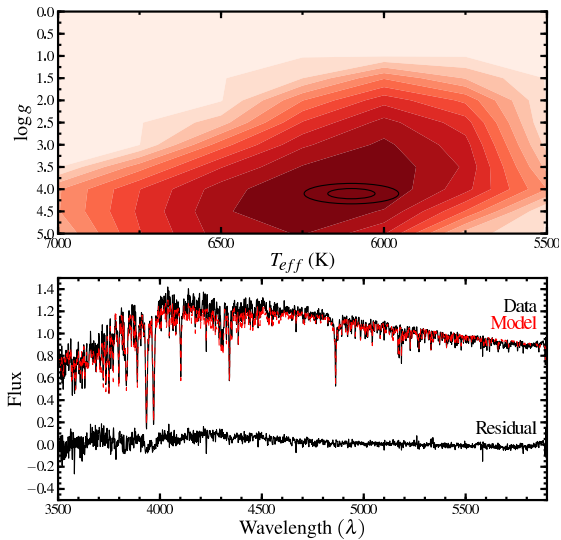}
  \caption{
    Low resolution spectral classification of \hatcur{} using
    WiFeS. Top: Contours marking the \teff{} --
    \logg{} probability space are shown. Typical 1 and 2$\sigma$
    uncertainties for WiFeS-derived stellar parameters are shown. Bottom: The observed
    spectrum (black solid) is plotted against the best fitting template
    spectrum (red dashed), with residuals plotted beneath.}
  \label{fig:wifes_spectype}
\end{figure}

The stellar properties \teff, \logg, \feh, and interstellar extinction
(E(B-V)) are derived via a grid search, minimising the \chisq{} between
the observed spectrum and synthetic templates from the MARCS model
atmospheres \citep{2008A&amp;A...486..951G}. The search intervals are
250\,K in \teff, 0.5 dex in \logg{} and \feh. A restricted \teff{} search
space is established using 2MASS J-K colours. Extinction is applied
according to \citet{1989ApJ...345..245C}, with E(B-V)
values ranging from 0 to the maximum extinction from the
\citet{1998ApJ...500..525S} maps. The \teff{} -- \logg{} probability space
for HATS-3 is plotted in Figure~\ref{fig:wifes_spectype}, along with
observed spectrum and best fitting template.

Since the differentiation of giants and dwarfs is of particular
importance, we place more weight on the \logg{} sensitive spectral
features during the \chisq{} calculations. These regions include the MgH
feature \citep[e.g.][]{1985MNRAS.212..497B,1994AstL...20..755B}, the
Mg b triplet \citep[e.g.][]{1997AJ....113.1865I} for cooler stars, and
the Balmer jump for hotter stars \citep[e.g.][]{2007PASP..119..605B}.

\begin{figure}[ht!]
\plotone{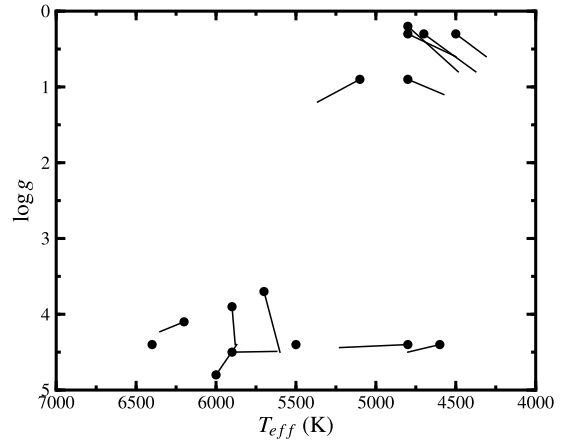}
\caption{ Benchmarking the WiFeS spectral classifications.  \teffstar
  and \logg\ values derived from our WiFeS spectral classification are
  marked by solid points, the vectors point towards the corresponding
  published values as set out in Table~\ref{tab:wifesbenchmark}.}
  \label{fig:spectype_comparison}
\end{figure}

To estimate the uncertainties of the HATSouth \hatrecon{} spectral
typing, we observed nine planet-hosting stars with published stellar
parameters derived from high resolution spectroscopy (including the
\hatcur{} properties presented in this study). The range in \teffstar{},
\logg{}, \feh{} and V magnitude of these stars closely resemble the
HATSouth candidates, so they serve as good bench markers. In addition,
we also include six evolved stars with published stellar parameters from
high resolution spectroscopy \citep{2006AA...458..895D}. These stars
have low \logg{} values and demonstrate our ability to distinguish
between giants and dwarfs. The results are presented in
Table~\ref{tab:wifesbenchmark} and
Figure~\ref{fig:spectype_comparison}. The root-mean-square deviation
between the parameters derived from our WiFeS observations of the dwarf stars
and their published parameters are \teffstar = 200\,K, \logg = 0.35, and
\feh{} = 0.44.

WiFeS low-resolution \hatrecon{} spectral classification revealed that
\hatcur{} is an \hatcurISOspec\ dwarf with $\teffstar=6200\pm 200\,K$,
$\loggstar=4.1\pm0.4$, and $\feh=-0.5\pm0.4$.  These values are consistent with the
more precise values obtained from subsequent high resolution
spectroscopy (subsection \ref{sec:highrspec}).

\begin{figure}[ht!]
\plotone{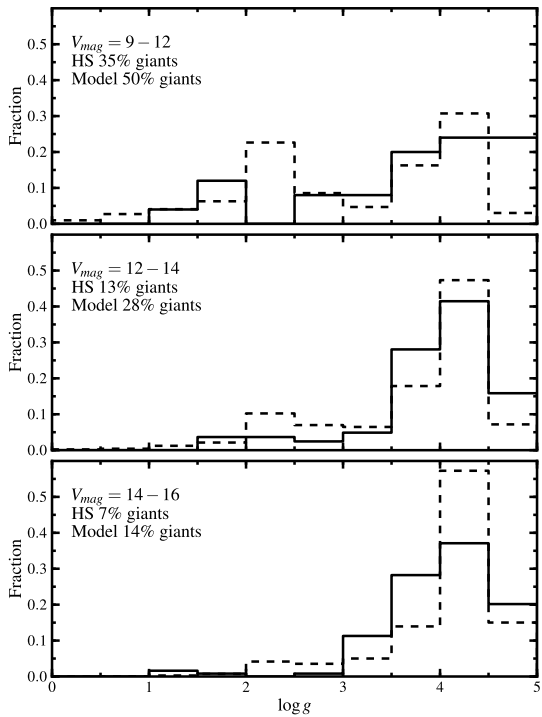}
  \caption{
    Histograms showing the \logg{} distribution for HATSouth candidates
    over different brightness bins.  The candidate population is plotted
    by solid lines, the Besan\c{c}on model population
    \citep{2003A&amp;A...409..523R} is plotted in dashed lines for
    reference. The contribution by giants is noted in each brightness
    bin.}
  \label{fig:logg_hist}
\end{figure}

To gauge the level of contamination to HATSouth candidates by false
positive scenarios involving giants, Figure~\ref{fig:logg_hist} plots
the histogram of the measured \logg{} value for all candidates
followed-up to date. We find that overall 12\% of the 240 HATSouth
candidates spectral typed to date are identified as giants ($\log g <
3.5$) by our \hatrecon{} spectroscopy.  For the expected galactic
population, approximated using the Besan\c{c}on model
\citep{2003A&amp;A...409..523R}, within a field centred at $\alpha =
300^\circ$, $\delta = −25^\circ$ and extending to 50 kpc distance, the
giant occurrence rate is 24\%.  Our candidate identification system is
not sensitive to planets orbiting giants (as these transit signatures
are so diluted as to be undetectable), so any ``transit'' seen around a
giant will generally either be red-noise mimicking a transit signal or
scenarios in which a giant is belended with an eclipsing binary.  The
contribution of giants is greater for brighter candidates, as expected
from the model population (see Figure~\ref{fig:logg_hist}). We find 35\%
of HATSouth candidates with magnitudes $9 < V < 12$ are giants, making
\hatrecon{} spectral typing extremely valuable over this magnitude
range.

\ifthenelse{\boolean{emulateapj}}{
    \begin{deluxetable*}{llrrr}
}{
    \begin{deluxetable}{llrrr}
}
\tablewidth{0pc}
\tabletypesize{\scriptsize}
\tablecaption{
    WiFeS \hatreconcap{} Spectral Typing
    Performance \label{tab:wifesbenchmark}
}
\tablehead{
    \multicolumn{1}{c}{Star} &
    \multicolumn{1}{c}{Reference}          &
    \multicolumn{1}{c}{\teff{}\tablenotemark{a}} &
    \multicolumn{1}{c}{\logg{}\tablenotemark{a}} &
    \multicolumn{1}{c}{\feh{} \tablenotemark{a}}\\
}
\startdata
WASP-4     & \citet{2008ApJ...675L.113W} & 5500(5500$\pm$150) & 4.4 (4.45+0.016/-0.029) & 0.0 (0.00$\pm$0.20)  \\
WASP-5     & \citet{2008MNRAS.387L...4A} & 5900 (5880$\pm$150) & 3.9 (4.40+0.039/-0.048) & 0.5 (0.09$\pm$0.09)  \\
WASP-7     & \citet{2009ApJ...690L..89H} & 6400 (6400$\pm$100) & 4.4 (4.36+0.01/-0.047) & -0.5 (0.00$\pm$0.1)  \\
WASP-8     & \citet{2010AA...517L...1Q} & 5700 (5600$\pm$80) & 3.7 (4.50$\pm$0.1) & -0.5 (0.17$\pm$0.07)  \\
WASP-29    & \citet{2010ApJ...723L..60H} & 4600 (4800$\pm$150) & 4.4 (4.50$\pm$0.2) & 0.0 (0.11$\pm$0.14)  \\
WASP-46    & \citet{2012MNRAS.422.1988A} & 5900 (5620$\pm$160) & 4.5 (4.49$\pm$0.02) &0.0 (-0.37$\pm$0.13)  \\
HATS-1     & \citet{penev:2013:hats1}    & 6000 (5870$\pm$100) & 4.8 (4.40$\pm$0.08) & -0.5 (-0.06$\pm$0.12)  \\
HATS-2     & \citet{mohler-fischer:2013:hats2} & 4800 (5227$\pm$95) & 4.4 (4.44$\pm$0.12) & -0.5 (0.15$\pm$0.05)  \\
\hatcur{}  & This work. & 6200 (\hatcurSMEteff{}) & 4.1 (\hatcurSMElogg{}) & -0.5 (\hatcurSMEzfeh{})  \\
HD36702 & \citet{2006AA...458..895D} & 4800 (4485$\pm$111) & 0.2 (0.8$\pm$0.15) & -2.0 (-2.0$\pm$0.17)\\
HD29574 & \citet{2006AA...458..895D} & 4500 (4310$\pm$111) & 0.3 (0.6$\pm$0.15) & -2.0 (-1.9$\pm$0.17)\\
HD26297 & \citet{2006AA...458..895D} & 4800 (4500$\pm$111) & 0.3 (1.2$\pm$0.15) & -1.5 (-1.7$\pm$0.17)\\
HD20453 & \citet{2006AA...458..895D} & 5100 (5365$\pm$111) & 0.9 (1.2$\pm$0.15) & -1.5 (-2.0$\pm$0.17)\\
HD103036 & \citet{2006AA...458..895D} & 4700 (4375$\pm$111) & 0.3 (0.8$\pm$0.15) & -1.5 (-1.7$\pm$0.17)\\
HD122956 & \citet{2006AA...458..895D} & 4800 (4575$\pm$111) & 0.9 (1.1$\pm$0.15) & -2.0 (-1.8$\pm$0.17)\\
%
\enddata 
\tablenotetext{a}{
        Literature value given in parentheses.
}
\ifthenelse{\boolean{emulateapj}}{
    \end{deluxetable*}
}{
    \end{deluxetable}
}

\subsubsection{\hatreconcap{} WiFeS Radial Velocities}
\label{sec:wifesrv}
In addition to determining stellar parameters, we also use WiFeS on the
ANU 2.3 m telescope to look for radial velocity variations above
$\sim$2\,\kms. Such variations indicate the transiting body is typically a
stellar-mass object rather than an exoplanet, and effectively rules out
the candidate as a transiting exoplanet. The observations are timed to
phase quadratures, where the expected velocity difference is greatest.

For radial velocity measurements, we use the red arm of the WiFeS
spectrograph with the R7000 grating and RT480 dichroic. This results in
$R\equiv \lambda/\Delta \lambda = 7000$ over 5200--7000\,\AA.\@ Wavelength
solutions are provided by bracketing NeFeAr arc lamp exposures, with a
further first order correction made using telluric Oxygen B band lines
at 6882--6906\,\AA.\@ Radial velocities are derived via cross
correlation against RV standard stars \citep{2002ApJS..141..503N}
exposures taken every night. Further details of the observing set-up and
the data reduction pipeline can be found in \citep{penev:2013:hats1}.

Simultaneous radial velocities are also derived for any neighbours
within the $12\times38\arcsec$ WiFeS field of view. Significant
velocity variations for any close-neighbours consistent with the
photometric ephemeris are indicative of blended eclipsing binary
scenarios, and are subsequently rejected.

\begin{figure} [ht]
\plotone{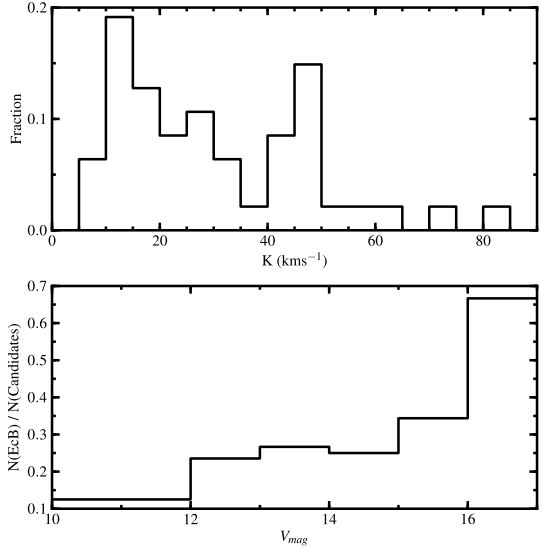}
\caption{
Top: Distribution of radial velocity semi-amplitude $(K)$ for the
stellar mass binaries identified from WiFeS medium resolution
measurements. Bottom: Fractional contamination by eclipsing binaries in
the HATSouth candidate population as a function of $V$ band magnitude.
\label{fig:ecb_hist}}
\end{figure}

To date 184 HATSouth candidates have been monitored using WiFeS
multi-epoch radial velocities measurements with enough phase coverage to
constrain any velocity variation at the 2\,\kms\ level.  We find 51
(27\%) to be stellar mass binaries.  Figure~\ref{fig:ecb_hist} presents
the distribution of the radial velocity orbit semi-amplitude $(K)$ of
these candidates. We find two peaks in the velocity amplitude, one at
$\sim 10$\,\kms, representing the F-M stellar binary population, and
another at $\sim 45$\,\kms, representing a larger mass ratio population
that exhibits shallow grazing transits. In both scenarios, the resulting
transit depth is similar to that of a planet-star system, and are
identified as potential planetary systems from their discovery
lightcurve.  Figure~\ref{fig:ecb_hist} also presents the fraction of
eclipsing stellar mass systems within our sample as a function of
candidate $V$-band magnitude.  We find a paucity of eclipsing binaries
for the brighter candidates.  The reason for this is that these
lightcurves are of higher precision that allows better determination of
the shape of the transit feature.  Also, as was pointed out in
Section~\ref{sec:wifesspec}, a higher fraction of the candidates are
giants.  There is an excess of eclipsing binaries for the faintest
candidates, for which only deep transits can be identified from the
lower precision discovery lightcurves.  The large number of
contaminating eclipsing binaries for the faintest candidates highlights
the particular advantage of medium resolution observations in this
regime.  Although faint, these candidates are of interest to the
HATSouth survey as they contain a high fraction of low mass stars that
are not easily probed by other wide-field transit surveys.

We obtained one 500\,s exposure on each of three consecutive nights from
2012 April 10 -- 12 to measure the radial velocity for \hatcur{} over
its full phase.  The radial velocities were clustered within 1\,\kms of
each other on each night, indicating that the transiting body could not
be of stellar mass.

\subsection{High Resolution Spectroscopy}
\label{sec:highrspec}
High-resolution spectroscopy is only carried out on candidates that
pass the screening process involved with the \hatrecon{} spectroscopy set
out in Section \ref{sec:reconspec}.  This allows us to focus our
time-intensive high-resolution spectroscopy on the targets that are most
likely to host planets.

\begin{figure} [ht]
\plotone{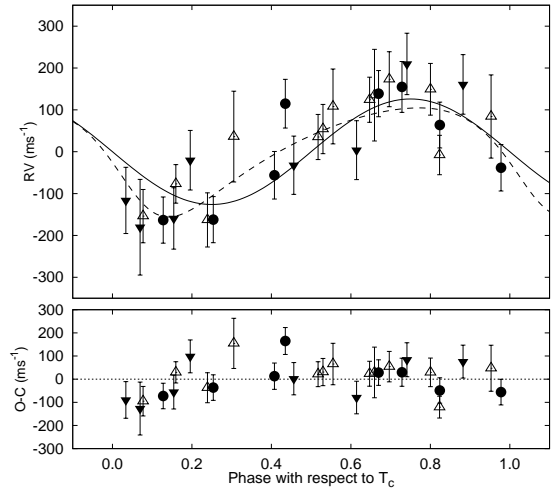}
\caption{
    {\em Top panel:} Radial velocity measurements for
    \hbox{\hatcur{}} from CORALIE (dark filled circles), FEROS (open
    triangles) and CYCLOPS (filled triangles) shown as a function of
    orbital phase, together with our best-fit circular (solid line) and
    eccentric (dotted line) models.  Zero phase corresponds to the time
    of mid-transit.  The center-of-mass velocity has been subtracted. 
    {\em Second panel:} Velocity $O\!-\!C$ residuals from the best fit
    circular model.  The error bars include a component from
    astrophysical/instrumental jitter allowed to differ for the three
    instruments.  
\label{fig:rv}}
\end{figure}

\hatcur{} was monitored by three different echelle spectrographs capable
of measuring high-precision radial velocities over the period 2012 April
-- June.  \reftabl{specobssummary} summarizes these high-resolution
spectroscopic observations.  Thirteen observations were taken with FEROS
\citep{kaufer:1998} on the MPG/ESO2.2\,m telescope at La Silla
Observatory, Chile.  A further eight observations of \hatcur{} were
taken by both CORALIE \citep{2000A&A...354...99Q} on the Swiss Leonard
Euler 1\,m telescope at La Silla Observatory, Chile, and UCLES (using
the CYCLOPS fibre feed) on the 3.9\,m AAT at Siding Spring Observatory,
Australia.  For a description of the observations and data
reduction employed for these instruments, see \citet{penev:2013:hats1}.
The radial velocity measurements for these observations are set out in
Table \ref{tab:rvs}, and are plotted after phase-wrapping to the
best-fit period (see Section \ref{sec:analysis}) in Figure \ref{fig:rv}.
Where possible we calculated the bisector spans of the cross-correlation
functions, and they did not vary in phase with the radial velocity
measurements.

\ifthenelse{\boolean{emulateapj}}{
    \begin{deluxetable*}{llrrrrr}
}{
    \begin{deluxetable}{llrrrrr}
}
\tablewidth{0pc}
\tabletypesize{\scriptsize}
\tablecaption{
    Summary of spectroscopic observations \label{tab:specobssummary}
}
\tablehead{
    \multicolumn{1}{c}{Telescope/} &
    \multicolumn{1}{c}{Date}          &
    \multicolumn{1}{c}{Number of} &
    \multicolumn{1}{c}{Exposure} &
    \multicolumn{1}{c}{Resolution}          &
    \multicolumn{1}{c}{SNR\tablenotemark{a}}&
    \multicolumn{1}{c}{Wavelength}\\
    \multicolumn{1}{c}{Instrument} &
    \multicolumn{1}{c}{Range}          &
    \multicolumn{1}{c}{Observations} &
    \multicolumn{1}{c}{Times (s)} &
    \multicolumn{1}{c}{}          &
    \multicolumn{1}{c}{}&
    \multicolumn{1}{c}{coverage (\AA)}\\
}
\startdata
\bf{\hatreconcap{}}\\
ANU 2.3\,m/WiFeS     & 2012 Apr 10 & 1 & 300 &3000  & 100   & 3500--6000 \\
ANU 2.3\,m/WiFeS     & 2012 Apr 10--12 & 3 & 500 & 7000  & 50 & 5200--7000 \\
\\
\bf{High Precision Radial Velocity}\\
AAT 3.9\,m/CYCLOPS   & 2012 May 5--11 & 8 & 1500 & 70000 & 20  & 4540--7340\\
Euler 1.2\,m/Coralie & 2012 Jun 2--7 & 8 & 1800 & 60000 & 20 & 3850--6900 \\
MPG/ESO 2.2\,m/FEROS & 2012 Apr 1--Jun 8 & 13 & 2700 & 48000 & 20 & 3500--9200 \\

%
\enddata 
\tablenotetext{a}{
        The approximate signal-to-noise per resolution element.
}
\ifthenelse{\boolean{emulateapj}}{
    \end{deluxetable*}
}{
    \end{deluxetable}
}

\ifthenelse{\boolean{emulateapj}}{
    \begin{deluxetable*}{lrrrr}
}{
    \begin{deluxetable}{lrrrr}
}
\tablewidth{0pc}
\tablecaption{
    Relative radial velocity measurements of
    \hatcur{}.
    \label{tab:rvs}
}
\tablehead{
    \colhead{BJD} & 
    \colhead{RV\tablenotemark{a}} & 
    \colhead{\ensuremath{\sigma_{\rm RV}}\tablenotemark{b}} & 
        \colhead{Phase} &
        \colhead{Instrument}\\
    \colhead{\hbox{(2\,454\,000$+$)}} & 
    \colhead{(\ms)} & 
    \colhead{(\ms)} &
        \colhead{} &
        \colhead{}
}
\startdata
$ 2019.89700 $ & $   124 $ & $    53 $  & $   0.647 $ & FEROS \\
$ 2032.87600 $ & $    36 $ & $   108 $  & $   0.305 $ & FEROS \\
$ 2035.90600 $ & $   -77 $ & $    45 $  & $   0.159 $ & FEROS \\
$ 2036.88374 $ & $   115 $ & $    28 $  & $   0.435 $ & Coralie \\
$ 2037.92527 $ & $   155 $ & $    33 $  & $   0.729 $ & Coralie \\
$ 2040.85700 $ & $   108 $ & $    89 $  & $   0.555 $ & FEROS \\
$ 2053.19936 $ & $  -116 $ & $    45 $  & $   0.034 $ & AAT \\
$ 2053.32636 $ & $  -180 $ & $    94 $  & $   0.070 $ & AAT \\
$ 2055.25949 $ & $     4 $ & $    27 $  & $   0.614 $ & AAT \\
$ 2056.20988 $ & $   161 $ & $    29 $  & $   0.882 $ & AAT \\
$ 2057.17510 $ & $  -159 $ & $    34 $  & $   0.154 $ & AAT \\
$ 2057.32183 $ & $   -20 $ & $    29 $  & $   0.196 $ & AAT \\
$ 2058.24637 $ & $   -32 $ & $    25 $  & $   0.456 $ & AAT \\
$ 2059.25685 $ & $   210 $ & $    34 $  & $   0.741 $ & AAT \\
$ 2076.83800 $ & $   173 $ & $    65 $  & $   0.697 $ & FEROS \\
$ 2077.74700 $ & $    84 $ & $    99 $  & $   0.953 $ & FEROS \\
$ 2078.76300 $ & $  -163 $ & $    64 $  & $   0.239 $ & FEROS \\
$ 2079.79400 $ & $    54 $ & $    58 $  & $   0.530 $ & FEROS \\
$ 2080.75400 $ & $   149 $ & $    61 $  & $   0.800 $ & FEROS \\
$ 2080.83500 $ & $    -8 $ & $    46 $  & $   0.823 $ & FEROS \\
$ 2080.83735 $ & $    64 $ & $    20 $  & $   0.824 $ & Coralie \\
$ 2081.73600 $ & $  -154 $ & $    63 $  & $   0.077 $ & FEROS \\
$ 2081.91580 $ & $  -163 $ & $    21 $  & $   0.128 $ & Coralie \\
$ 2082.90893 $ & $   -56 $ & $    25 $  & $   0.408 $ & Coralie \\
$ 2083.80200 $ & $   135 $ & $   109 $  & $   0.659 $ & FEROS \\
$ 2083.83791 $ & $   139 $ & $    21 $  & $   0.670 $ & Coralie \\
$ 2084.93164 $ & $   -38 $ & $    22 $  & $   0.978 $ & Coralie \\
$ 2085.91066 $ & $  -162 $ & $    21 $  & $   0.254 $ & Coralie \\
$ 2086.84600 $ & $    35 $ & $    53 $  & $   0.517 $ & FEROS \\
    [-1.5ex]
\enddata
\tablenotetext{a}{
        The zero-point of these velocities is arbitrary. An overall
        offset $\gamma_{\rm rel}$ fitted separately to the CORALIE,
        FEROS and CYCLOPS velocities in \refsecl{analysis} has been
        subtracted.
}
\tablenotetext{b}{
        Internal errors excluding the component of
        astrophysical/instrumental jitter considered in
        \refsecl{analysis}.
}
\ifthenelse{\boolean{emulateapj}}{
    \end{deluxetable*}
}{
    \end{deluxetable}
}

\subsection{Photometric follow-up observations}
\label{sec:phot}

High-precision photometric follow-up is important to determining the
precise orbital parameters of the exoplanet system and the planetary
radius.  We used two facilities for this task: the Spectral camera on the
2.0\,m Faulkes Telescope South (FTS) and the GROND camera on the
MPG/ESO 2.2\,m telescope.  A summary of the high precision photometric
follow-up is presented in Table~\ref{tab:photobs}.

\subsubsection{FTS~2\,m/Spectral}
\label{sec:FTS}
FTS is a fully automated, robotic telescope operated as part of the
Las Cumbres Observatory Global Telescope (LCOGT) Network.  The queue-based
scheduling allows for transiting planets to be easily monitored at the
time of the transit event.  The ``Spectral'' camera with an \band{i}
filter is employed for our transit observations of \hatcurb{}.  The
camera has a 4K$\times$4K array of $0.15^{\prime\prime}$ pixels, and we use it
with $2\times2$ binning to reduce readout time.  The telescope is
slightly defocused to reduce the effect of imperfect flat-fielding and
to allow for slightly longer exposure times without saturating.  For
\hatcur{} the exposures were 30\,s which provided for 50\,s cadence
photometry given the CCD readout time of the Spectral
camera.  The raw fits files are reduced automatically via the LCOGT
reduction pipeline, which includes flat-field correction and fitting an
astrometric solution.  Photometry is performed on the reduced images
using an automated pipeline based on aperture photometry with Source
Extractor \citep{bertin:1996}.

On the night of 2012 June 20, we monitored \hatcur{} with FTS.
Although the observations only started mid-ingress and finished before
the egress began, this information allowed us to
update the period and phase, and alert us to the need for further
follow-up photometry.  On 2012 July 15 we again monitored \hatcur{} with
FTS, this time covering the entire transit event.  This \lc{} is presented
in Figure \ref{fig:lc}.

\subsubsection{ESO~2.2\,m/GROND}
\label{sec:grond}
GROND is a seven-channel imager on the MPG/ESO 2.2\,m telescope at La
Silla Observatory in Chile \citep{2008PASP..120..405G}.  Though it is
primarily designed for rapid observations of gamma-ray burst afterglows,
it has proved a very useful instrument for multi-band, high precision
follow-up \lcs{} for transiting planets in general
\citep[e.g.][]{2013MNRAS.430.2932M}, and in particular HATSouth planet
discoveries \citep[e.g.][]{penev:2013:hats1,mohler-fischer:2013:hats2}.

\begin{figure}[!ht]
\plotone{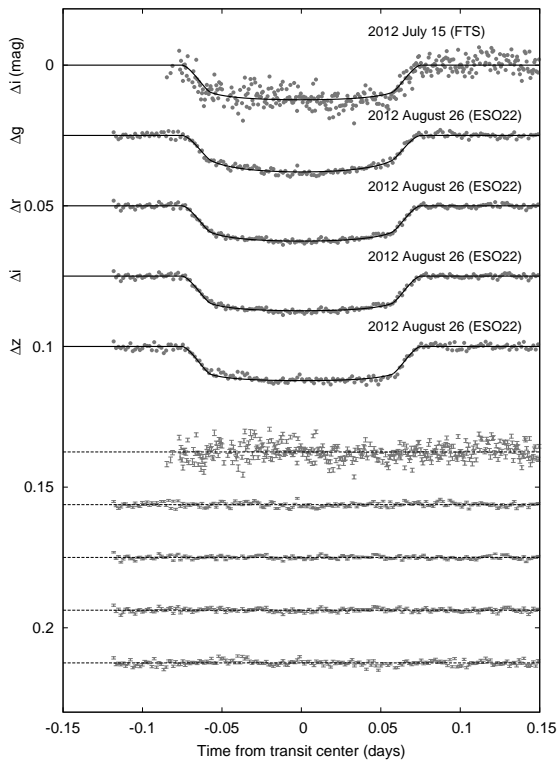}
\caption{
        Unbinned instrumental Sloan $g$-, $r$-, $i$- and \band{z}
        transit \lcs{} of \hatcur{}.  The dates and instruments used
        for each event are indicated.  The \lcs{} have been
        detrended using the EPD and TFA processes.  \Lcs{} are shifted for clarity.  Our best fit is shown by the
        solid lines.  Residuals from the fits are displayed at the
        bottom, in the same order as the top \lcs.
\label{fig:lc}} \end{figure}

On 2012 August 26 we used GROND to simultaneously monitor an entire
transit of \hatcurb{} in four bands, similar to Sloan $g$-, $r$-, $i$-
and \band{z}).  The telescope was defocused, and the exposure time was
set to 62\,s in all bands, which resulted in an effective cadence of
129\,s.  The data were reduced in the standard manner, and aperture
photometry was performed using an IDL/ATROLIB implemtation of \daophot{} \citep{1987PASP...99..191S, 2009MNRAS.396.1023S} .  The \lcs{} for each of the four bands are set out
in Figure \ref{fig:lc}.

\ifthenelse{\boolean{emulateapj}}{
        \begin{deluxetable*}{lrrrrr} }{
        \begin{deluxetable}{lrrrrr} 
    }
        \tablewidth{0pc}
        \tablecaption{Differential photometry of
        \hatcur\label{tab:phfu}} \tablehead{ \colhead{BJD} &
        \colhead{Mag\tablenotemark{a}} &
        \colhead{\ensuremath{\sigma_{\rm Mag}}} &
        \colhead{Mag(orig)\tablenotemark{b}} & \colhead{Filter} &
        \colhead{Instrument} \\ \colhead{\hbox{~~~~(2\,400\,000$+$)~~~~}}
        & \colhead{} & \colhead{} & \colhead{} & \colhead{} &
        \colhead{} } \startdata 
$ 55100.48159 $ & $  -0.00404 $ & $   0.00277 $ & $ \cdots $ & $ r$ &         HS\\
$ 55366.57071 $ & $   0.00091 $ & $   0.00242 $ & $ \cdots $ & $ r$ &         HS\\
$ 55334.64007 $ & $   0.01110 $ & $   0.00268 $ & $ \cdots $ & $ r$ &         HS\\
$ 55359.47536 $ & $   0.00753 $ & $   0.00245 $ & $ \cdots $ & $ r$ &         HS\\
$ 55391.40621 $ & $  -0.00212 $ & $   0.00256 $ & $ \cdots $ & $ r$ &         HS\\
$ 55380.76328 $ & $  -0.00029 $ & $   0.00250 $ & $ \cdots $ & $ r$ &         HS\\
$ 55437.52939 $ & $  -0.00387 $ & $   0.00252 $ & $ \cdots $ & $ r$ &         HS\\
$ 55423.33825 $ & $   0.00704 $ & $   0.00250 $ & $ \cdots $ & $ r$ &         HS\\
$ 55437.52985 $ & $   0.00690 $ & $   0.00425 $ & $ \cdots $ & $ r$ &         HS\\
$ 55093.38841 $ & $   0.00932 $ & $   0.00260 $ & $ \cdots $ & $ r$ &         HS\\
        [-1.5ex]
\enddata \tablenotetext{a}{
     The out-of-transit level has been subtracted. For the
     HATSouth \lc{} (rows with ``HS'' in the
     Instrument column), these magnitudes have been
     detrended using the EPD and TFA procedures prior to
     fitting a transit model to the \lc. Primarily
     as a result of this detrending, but also due to
     blending from neighbors, the apparent HATSouth transit
     depth is $\sim 86\%$ that of the true depth in the
     Sloan~$r$ filter. For the follow-up \lcs{} (rows
     with an Instrument other than ``HS'') these magnitudes
     have been detrended with the EPD and TFA procedures,
     carried out simultaneously with the transit fit (the
     transit shape is preserved in this process).
}
\tablenotetext{b}{
        Raw magnitude values without application of the EPD and TFA
        procedures.  This is only reported for the follow-up \lcs.
}
\tablecomments{
        This table is available in a machine-readable form in the
        online journal.  A portion is shown here for guidance regarding
        its form and content.
} \ifthenelse{\boolean{emulateapj}}{ \end{deluxetable*} }{ \end{deluxetable} }

\section{Analysis}
\label{sec:analysis}

\subsection{Properties of the host star \hatcur{}}
\label{sec:stelparam}

The stellar parameters of \hatcur{}, including \teffstar{}, \loggstar{},
\feh{}, and $\vsini$ are derived from the high-resolution FEROS spectra.
In a procedure similar to that presented in
\citet{mohler-fischer:2013:hats2}, the thirteen spectra were analysed using the ``Spectroscopy Made Easy''
software package \citep[SME;][]{valenti:1996}.  The stellar parameters we
list are the mean of the values derived from each spectrum, weighted by
the signal-to-noise of the spectrum.  The uncertainties in the stellar
parameters are derived from the distribution of these parameters over
the thirteen spectra.

\begin{figure}[!ht]
\plotone{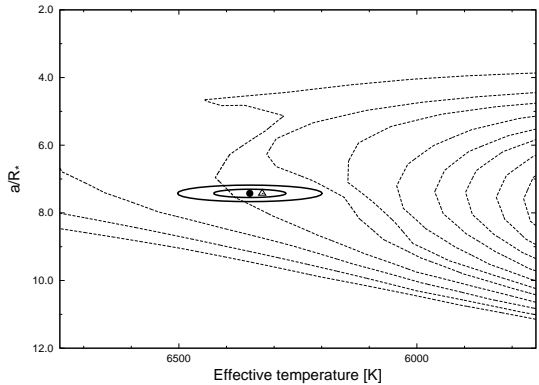}
\caption{
    Model isochrones from \cite{\hatcurisocite} for the measured
    metallicity of \hatcur, \feh = \hatcurSMEiizfehshort, and ages of
    0.2\,Gyr (lowermost line), and 1 to 10\,Gyr in 1\,Gyr increments
    (left to right).  The adopted values of $\teffstar$ and \arstar\ are
    shown together with their 1$\sigma$ and 2$\sigma$ confidence
    ellipsoids.  The open triangle shows the values from the initial SME
    iteration.
    \label{fig:iso}}
\end{figure}

Following the method described in \citet{sozzetti:2007} and applied in
\citet{penev:2013:hats1}, we determine fundamental stellar properties
(mass, radius, age, and luminosity) based on the mean stellar density
derived from the \lc{} fitting, the \teffstar{} from the SME analysis,
and the Yonsei-Yale \citep[Y2;][]{yi:2001} stellar evolution models.
This analysis provides a value of \loggstar=4.23$\pm$0.01, which is more
precise than can be obtained via SME alone.  We then fix this as the
\loggstar{} for \hatcur{} and repeat the SME analysis to determine the
final stellar parameters listed in Table~\ref{tab:stellar}.  The
1$\sigma$ and 2$\sigma$ confidence ellipsoids in \teffstar{} and
\arstar\ are plotted in Figure \ref{fig:iso}, along with the Y2
isochrones for the SME determined \feh, and a range of stellar ages.  We
find that \hatcur{} is an \hatcurISOspec\ dwarf host, the first for the
HATSouth survey, with \teffstar=\hatcurSMEteff\,K.  

\subsection{Excluding Blends}
\label{sec:blends}

To rule out the possibility that \hatcur{} is a blended stellar binary
system that mimics the observable properties of a transiting planet
system we conducted a blend analysis following the procedure described
in \citet{hartman:2011:hat3233}. This procedure involves modelling the
photometric light curves, photometry taken from public catalogs and
calibrated to an absolute scale, and spectroscopically determined
stellar atmospheric parameters.  We compare the fits from a model
consisting of a single planet transiting a star, to models of blended
stellar systems with components having properties constrained by stellar
evolution models.

We find that the single-star plus transiting
planet model fits the data better than any non-planetary blend
scenario. As is typically the case, there are some scenarios, in which
the two brightest components in the blend are of similar mass, that we
cannot rule out with greater than 5$\sigma$ confidence based on the
photometry and atmospheric parameters. However, in all such cases
there would be a secondary component in the blended object with a
brightness that is $>38\%$ that of the primary star. That secondary
component would itself be undergoing an orbit with a semiamplitude of
several tens of km/s. Such a blend would have been easily detected,
either as a double-lined spectrum, or via several \kms\ bisector-span
variations.

\subsection{Global Modeling of Data}
\label{sec:global}
In order to determine the physical planetary parameters, we carried out
a joint, Markov-Chain Monte Carlo modelling of the HATSouth photometry,
follow-up photometry, and the radial velocity measurements.  The
methodology is described fully in \citet{bakos:2010:hat11}.  The resulting planetary parameters are set
out in Table \ref{tab:planetparam}.  We list two sets of planet
parameters: for the case of a circular orbit ($e=0$) and for the case
that eccentricity is allowed to vary, whereby we get a best fit
eccentricity of $e=0.25\pm0.10$.

To estimate the significance of the non-zero eccentricity measurement we
find that a \citet{lucy:1971} test gives a non-negligible (7\%)
probability that the RV observations are consistent with a circular
orbit, while the free-eccentricity and fixed-circular-orbit models yield
nearly identical values for the Bayesian Information Criterion. We
conclude that we cannot rule out a circular orbit based on the current
RV observations.


\begin{deluxetable*}{lrrl}
\tablewidth{0pc}
\tabletypesize{\scriptsize}
\tablecaption{
    Stellar parameters for \hatcur{}
    \label{tab:stellar}
}
\tablehead{
    \multicolumn{1}{c}{~~~~~~~~Parameter~~~~~~~~}   &
    \multicolumn{1}{c}{Value} &
    \multicolumn{1}{c}{Value} &
    \multicolumn{1}{c}{Source} \\
    \multicolumn{1}{c}{}    &
    \multicolumn{1}{c}{\textbf{Circular}} &
    \multicolumn{1}{c}{Eccentric} &
    \multicolumn{1}{c}{}
}
\startdata
\noalign{\vskip -3pt}
\sidehead{Spectroscopic properties}
~~~~$\teffstar$ (K)\dotfill         &  \hatcurSMEteff &  \hatcurSMEteffeccen       & SME\tablenotemark{a}\\
~~~~$\feh$\dotfill                  &  \hatcurSMEzfeh &  \hatcurSMEzfeheccen       & SME                 \\
~~~~$\vsini$ (\kms)\dotfill         &  \hatcurSMEvsin &  \hatcurSMEvsineccen       & SME                 \\
\sidehead{Photometric properties}
~~~~$V$ (mag)\dotfill               &  \hatcurCCapassmV &  \hatcurCCapassmVeccen      & APASS                \\
~~~~$B$ (mag)\dotfill               &  \hatcurCCapassmB &  \hatcurCCapassmBeccen      & APASS                \\
~~~~$J$ (mag)\dotfill               &  \hatcurCCtwomassJmag &  \hatcurCCtwomassJmageccen & 2MASS           \\
~~~~$H$ (mag)\dotfill               &  \hatcurCCtwomassHmag &  \hatcurCCtwomassHmageccen & 2MASS           \\
~~~~$K_s$ (mag)\dotfill             &  \hatcurCCtwomassKmag &  \hatcurCCtwomassKmageccen & 2MASS           \\
\sidehead{Derived properties}
~~~~$\mstar$ ($\msun$)\dotfill      &  \hatcurISOmlong &  \hatcurISOmlongeccen      & \hatcurisoshort+\hatcurlumind+SME \tablenotemark{b}\\
~~~~$\rstar$ ($\rsun$)\dotfill      &  \hatcurISOrlong &  \hatcurISOrlongeccen      & \hatcurisoshort+\hatcurlumind+SME         \\
~~~~$\loggstar$ (cgs)\dotfill       &  \hatcurISOlogg &  \hatcurISOloggeccen       & \hatcurisoshort+\hatcurlumind+SME         \\
~~~~$\lstar$ ($\lsun$)\dotfill      &  \hatcurISOlum &  \hatcurISOlumeccen        & \hatcurisoshort+\hatcurlumind+SME         \\
~~~~$M_V$ (mag)\dotfill             &  \hatcurISOmv &  \hatcurISOmveccen         & \hatcurisoshort+\hatcurlumind+SME         \\
~~~~$M_K$ (mag,\hatcurjhkfilset)\dotfill &  \hatcurISOMK &  \hatcurISOMKeccen    & \hatcurisoshort+\hatcurlumind+SME         \\
~~~~Age (Gyr)\dotfill               &  \hatcurISOage  &  \hatcurISOageeccen        & \hatcurisoshort+\hatcurlumind+SME         \\
~~~~Distance (pc)\dotfill           &  \hatcurXdist   &  \hatcurXdisteccen         & \hatcurisoshort+\hatcurlumind+SME\\
~~~~E(B-V) \dotfill           &  \hatcurXEBV   &  \hatcurXEBVeccen         & \hatcurisoshort+\hatcurlumind+SME\\
[-1.5ex]
\enddata
\tablenotetext{a}{
    SME = ``Spectroscopy Made Easy'' package for the analysis of
    high-resolution spectra \citep{valenti:1996}.  These parameters
    rely primarily on SME, but have a small dependence also on the
    iterative analysis incorporating the isochrone search and global
    modeling of the data, as described in the text.
}
\tablenotetext{b}{
    \hatcurisoshort+\hatcurlumind+SME = Based on the \hatcurisoshort\
    isochrones \citep{\hatcurisocite}, \hatcurlumind\ as a luminosity
    indicator, and the SME results.
}
\end{deluxetable*}

\begin{deluxetable*}{lrr}
\tabletypesize{\scriptsize}
\tablecaption{Orbital and planetary parameters\label{tab:planetparam}}
\tablehead{
        \multicolumn{1}{c}{~~~~~~~~~~~~~~~Parameter~~~~~~~~~~~~~~~} &
        \multicolumn{1}{c}{Value} &
        \multicolumn{1}{c}{Value} \\
        \multicolumn{1}{c}{} &
        \multicolumn{1}{c}{\textbf{Circular}} &
        \multicolumn{1}{c}{Eccentric} 
}
\startdata
\noalign{\vskip -3pt}
\sidehead{\Lc{} parameters}
~~~$P$ (days)             \dotfill    & $\hatcurLCP$ & $\hatcurLCPeccen$    \\
~~~$T_c$ (${\rm BJD}$)    
      \tablenotemark{a}   \dotfill    & $\hatcurLCT$ & $\hatcurLCTeccen$    \\
~~~$T_{14}$ (days)
      \tablenotemark{a}   \dotfill    & $\hatcurLCdur$ & $\hatcurLCdureccen$ \\
~~~$T_{12} = T_{34}$ (days)
      \tablenotemark{a}   \dotfill    & $\hatcurLCingdur$ & $\hatcurLCingdureccen$ \\
~~~$\arstar$              \dotfill    & $\hatcurPPar$ & $\hatcurPPareccen$ \\
~~~$\zrstar$\tablenotemark{b}              \dotfill    & $\hatcurLCzeta$ & $\hatcurLCzetaeccen$ \\
~~~$\rpl/\rstar$          \dotfill    & $\hatcurLCrprstar$ & $\hatcurLCrprstareccen$ \\
~~~$b \equiv a \cos i/\rstar$
                          \dotfill    & $\hatcurLCimp$ & $\hatcurLCimpeccen$  \\
~~~$i$ (deg)              \dotfill    & $\hatcurPPi$ & $\hatcurPPieccen$  \\

\sidehead{Limb-darkening coefficients \tablenotemark{c}}
~~~$a_g$ (linear term)   \dotfill    & $\hatcurLBig$ & $\hatcurLBigeccen$ \\
~~~$b_g$ (quadratic term) \dotfill    & $\hatcurLBiig$ & $\hatcurLBiigeccen$ \\
~~~$a_r$                 \dotfill    & $\hatcurLBir$ & $\hatcurLBireccen$ \\
~~~$b_r$                 \dotfill    & $\hatcurLBiir$ & $\hatcurLBiireccen$ \\
~~~$a_i$                 \dotfill    & $\hatcurLBii$ & $\hatcurLBiieccen$ \\
~~~$b_i$                  \dotfill    & $\hatcurLBiii$ & $\hatcurLBiiieccen$ \\

\sidehead{RV parameters}
~~~$K$ (\ms)              \dotfill    & $\hatcurRVK$ & $\hatcurRVKeccen$ \\
~~~$\sqrt{e}\cos\omega$ 
                          \dotfill    & $\cdots$ & $\hatcurRVrkeccen$ \\
~~~$\sqrt{e}\sin\omega$
                          \dotfill    & $\cdots$ & $\hatcurRVrheccen$ \\
~~~$e\cos\omega$ 
                          \dotfill    & $\cdots$ & $\hatcurRVkeccen$ \\
~~~$e\sin\omega$
                          \dotfill    & $\cdots$ & $\hatcurRVheccen$ \\
~~~$e$                    \dotfill    & $0$ & $\hatcurRVecceneccen$ \\
~~~$\omega$                    \dotfill    & $\cdots$ & $\hatcurRVomegaeccen$ \\
~~~CORALIE RV jitter (\ms)\tablenotemark{d}        
                          \dotfill    & $\hatcurRVjitterA$ & $\hatcurRVjitterAeccen$ \\
~~~FEROS RV jitter (\ms)        
                          \dotfill    & $\hatcurRVjitterB$ & $\hatcurRVjitterBeccen$ \\
~~~CYCLOPS RV jitter (\ms)
                          \dotfill    & $\hatcurRVjitterC$ & $\hatcurRVjitterCeccen$ \\

\sidehead{Planetary parameters}
~~~$\mpl$ ($\mjup$)       \dotfill    & $\hatcurPPmlong$ & $\hatcurPPmlongeccen$ \\
~~~$\rpl$ ($\rjup$)       \dotfill    & $\hatcurPPrlong$ & $\hatcurPPrlongeccen$ \\
~~~$C(\mpl,\rpl)$
    \tablenotemark{e}     \dotfill    & $\hatcurPPmrcorr$ & $\hatcurPPmrcorreccen$ \\
~~~$\rhopl$ (\gcmc)       \dotfill    & $\hatcurPPrho$ & $\hatcurPPrhoeccen$ \\
~~~$\log g_p$ (cgs)       \dotfill    & $\hatcurPPlogg$ & $\hatcurPPloggeccen$ \\
~~~$a$ (AU)               \dotfill    & $\hatcurPParel$ & $\hatcurPPareleccen$ \\
~~~$T_{\rm eq}$ (K)       \dotfill    & $\hatcurPPteff$ & $\hatcurPPteffeccen$ \\
~~~$\Theta$\tablenotemark{f}\dotfill  & $\hatcurPPtheta$ & $\hatcurPPthetaeccen$ \\
~~~$\langle F \rangle$ ($10^{\hatcurPPfluxavgdim}$\ergscmsq) 
\tablenotemark{g}         \dotfill    & $\hatcurPPfluxavg$ & $\hatcurPPfluxavgeccen$ \\
\enddata
\tablenotetext{a}{
    \ensuremath{T_c}: Reference epoch of mid transit that minimizes the
    correlation with the orbital period. BJD is calculated from UTC.
    \ensuremath{T_{14}}: total transit duration, time between first to
    last contact;
    \ensuremath{T_{12}=T_{34}}: ingress/egress time, time between first
    and second, or third and fourth contact.
}
\tablenotetext{b}{
    Reciprocal of the half duration of the transit used as a jump
    parameter in our MCMC analysis in place of $\arstar$. It is
    related to $\arstar$ by the expression $\zrstar = \arstar
    (2\pi(1+e\sin \omega))/(P \sqrt{1 - b^{2}}\sqrt{1-e^{2}})$
    \citep{bakos:2010:hat11}.
}
\tablenotetext{c}{
        Values for a quadratic law given separately for the Sloan~$g$,
        $r$, and $i$ filters.  These values were adopted from the
        tabulations by \cite{claret:2004} according to the
        spectroscopic (SME) parameters listed in \reftabl{stellar}.
}
\tablenotetext{d}{
    This jitter was added in quadrature to the RV uncertainties for
    each instrument such that \chisq/{\rm dof} = 1 for the
    observations from that instrument for the free eccentricity model.
}
\tablenotetext{e}{
    Correlation coefficient between the planetary mass \mpl\ and radius
    \rpl.
}
\tablenotetext{f}{
    The Safronov number is given by $\Theta = \frac{1}{2}(V_{\rm
    esc}/V_{\rm orb})^2 = (a/\rpl)(\mpl / \mstar )$
    \citep[see][]{hansen:2007}.
}
\tablenotetext{g}{
    Incoming flux per unit surface area, averaged over the orbit.
}
\end{deluxetable*}

\section{Discussion}
\label{sec:discussion}

\hatcurb{} is the third planet to be discovered as part of the ongoing
HATSouth survey for transiting exoplanets.  Although it is the least
massive of the three exoplanets with
$\mpl\approx\hatcurPPmshort$\,\mjup, the host star is the most massive
of the three host stars (\mstar=\hatcurISOmshort\,\msun).

When we allow the eccentricity to depart from e=0 in the global fit, we
find a high eccentricity (e=\hatcurRVecceneccen) for \hatcurb{}.
However like many planetary systems, the eccentricity is not well
constrained by the radial velocity measurements, and more data will be
required before this high eccentricity can be confirmed.

The $\vsini$=\hatcurSMEvsin{}\,\kms, combined with the large radius of
the planet ($\rpl=\hatcurPPrlong\,\rjup$) makes \hatcurb{} a prime
target for follow-up Rossiter-McLaughlin monitoring to determine the
spin-orbit alignment of the system.  Interestingly with
\teffstar=\hatcurSMEteff{}\,K this star lies just above the temperature
of 6250\,K which is noted as apparently marking a shift from aligned to
misaligned {\hj}s \citep{2010ApJ...718L.145W, 2012ApJ...757...18A}.

\begin{figure}[!ht]
\plotone{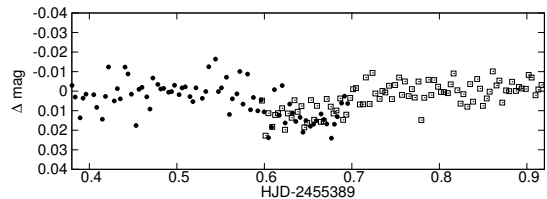}
\caption{
    A single transit event from 2010 July 11 simultaneously observed
    from the HATSouth telescopes in Namibia (filled circles) and Chile
    (open squares).  For two hours both telescopes were
    monitoring \hatcurb{} in-transit.  Such simultaneous monitoring is
    possible for much of the year with the HATSouth network.
\label{fig:overlap}}
\end{figure}

The power of the HATSouth network was illustrated in
\citet{penev:2013:hats1} whereby three consecutive transits were
detected at three different ground stations.  For \hatcurb{} we show
another aspect of the global network, which is that for much of the year
observations from the sites give considerable overlap.
Figure~\ref{fig:overlap} shows that in \textit{pre-discovery} a single
transit of \hatcurb{} was simultaneously monitored by two different
telescopes located over 8000\,km apart on the Earth.  This is testament
to an active, homogeneous, global network.

\begin{figure}[!ht]
\plotone{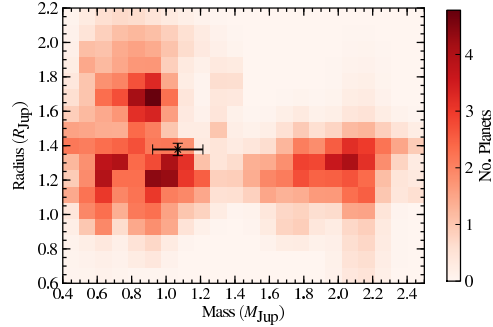}
\caption{
    A density plot showing the mass and radius for published transiting hot
    Jupiters, defined as planets with $\mpl>0.5\,\mjup$ and $P<10\,d$.
    \hatcurb{} is plotted in black with appropriate uncertainties.
    \label{fig:mass_radius}}
\end{figure}

There are 125 published transiting {\hj}s (defined as $\mpl>0.5\,\mjup$,
$P<10\,d$) with both radius and mass precisely determined\footnote{http://exoplanets.org}, so we can begin to plot the
mass--radius diagram in terms of a density distribution of known
exoplanets rather than just individual points.  Such a plot is presented
in Figure~\ref{fig:mass_radius}, with the position of \hatcurb{} marked
in black with appropriate error bars.  Interestingly, this diagram
reveals an under-density of planets at 1.2--1.6 \mjup, even though
there is no selection bias against detecting planets in this mass range
with transit surveys.  While it is possible that the gap is simply a
feature of small number statistics, it further supports the case for
expanding the sample of {\hj}s to probe global trends in the
populations.

\acknowledgements 

Development of the HATSouth project was funded by NSF MRI grant
NSF/AST-0723074, operations are supported by NASA grant NNX09AB29G, and
follow-up observations receive partial support from grant
NSF/AST-1108686.
Work at the Australian National University is supported by ARC Laureate
Fellowship Grant FL0992131.
Followup observations with the ESO~2.2\,m/FEROS instrument were
performed under MPI guaranteed time (P087.A-9014(A), P088.A-9008(A),
P089.A-9008(A)) and Chilean time (P087.C-0508(A)).
A.J.\ acknowledges support from Fondecyt project 1130857, Ministry of
Economy ICM Nucleus P10-022-F, Anillo ACT-086, and BASAL CATA PGB-06.
V.S.\ acknowledges support form BASAL CATA PFB-06.  M.R.\ acknowledges
support from a FONDECYT postdoctoral fellowship.  R.B.\ and N.E.\
acknowledge support from CONICYT-PCHA/Doctorado Nacional and Fondecyt
project 1130857.
This work is based on observations made with ESO Telescopes at the La
Silla Observatory under programme IDs P087.A-9014(A), P088.A-9008(A),
P089.A-9008(A), P087.C-0508(A), 089.A-9006(A), and
This paper also uses observations obtained with facilities of the Las
Cumbres Observatory Global Telescope.
Work at UNSW has been supported by ARC Australian Professorial Fellowship grant DP0774000,
ARC LIEF grant LE0989347 and ARC Super Science Fellowships FS100100046.
We acknowledge the use of the AAVSO Photometric All-Sky Survey (APASS),
funded by the Robert Martin Ayers Sciences Fund, and the SIMBAD
database, operated at CDS, Strasbourg, France.
Operations at the MPG/ESO 2.2\,m Telescope are jointly performed by the
Max Planck Gesellschaft and the European Southern Observatory.  The
imaging system GROND has been built by the high-energy group of MPE in
collaboration with the LSW Tautenburg and ESO\@.  We thank Timo Anguita
and R\'egis Lachaume for their technical assistance during the
observations at the MPG/ESO 2.2\,m Telescope.


\bibliographystyle{apj}
\bibliography{hatsbib}

\end{document}